# Bridging theory and experiment in virtually imaged phased array (VIPA) spectrometers

**Kiumars Aryana,[1,2,3] D. Michelle Bailey,[1] Solomon I. Woods,[2] and Adam J. Fleisher[1,*]**

[1]*Material Measurement Laboratory, National Institute of Standards and Technology, Gaithersburg, MD 20899, USA*
[2]*Physical Measurement Laboratory, National Institute of Standards and Technology, Gaithersburg, MD 20899, USA*
[3]*Department of Chemistry and Biochemistry, University of Maryland, College Park, MD 20742, USA*
**adam.fleisher@nist.gov*

**Abstract:** Virtually imaged phased array (VIPA) spectrometers provide high resolution and fast acquisition in a compact design, but their performance is sensitive to fabrication tolerances, component dimensions, and alignment. Here, leveraging numerical simulations validated by experimental data, we present a framework to identify the parameters that limit VIPA spectrometer resolution. This framework is applied to the construction of a new mid-infrared VIPA spectrometer, tested at wavelengths near $\lambda$ = 4.6 μm with both continuous-wave and frequency-comb laser sources, with a resolving power predicted by analytical expressions to be as high as $RP$ = 830 000 (corresponding to a resolution of $\delta\nu$ = 78 MHz). Validated numerical simulations, however, provided a more realistic estimate that captures limits set by all the optical components. By minimizing aberrations and optimizing alignment, a resolving power of $RP$ = 440 000 ($\delta\nu$ = 150 MHz) was experimentally achieved, corresponding to 80 % of the value predicted by numerical simulation of the entire spectrometer. The results bridge the gap between analytical expressions and experimental results for compact, high-resolution VIPA spectrometers to enable more efficient fabrication and advanced design across critical areas like space optics, line-by-line pulse shaping, and broadband spectral sensors.

## 1. Introduction

The unambiguous detection of multiple molecular species has applications in scientific research and industry, including for medical diagnostics [1–3] and imaging [4], chemical kinetics [5–7], plasma diagnostics [8–11], and astronomical discoveries [12,13]. Different spectroscopic techniques offer trade-offs in terms of their spectral range, resolution, measurement speed, and size. Fourier Transform Infrared (FTIR) spectroscopy is one of the most widely used techniques, noted for its broadband spectral coverage and commercial availability [14,15]. However, FTIR relies on moving components and its resolution is dependent upon the interferometer path length (with the exception being sub-nominal techniques [16]), a dimension which increases at longer wavelengths. Dispersive spectrometers are an alternative approach that also offer high resolving power, albeit over a narrower spectral range. Crossed dispersion spectrometers that leverage the high dispersion of a virtually imaged phased array (VIPA) employ two elements to spatially disperse broadband light onto a two-dimensional detector array. In these spectrometers, a VIPA and a diffraction grating together enable a simple and compact instrument design.

The VIPA, developed for wavelength demultiplexing [17–19], was first introduced in a cross-dispersed spectrometer for molecular fingerprinting in the visible at a wavelength of $\lambda$ = 633 nm [20]. That pioneering experiment achieved a resolving power of $RP$ = 400 000, based on a full-width at half-maximum (FWHM) definition [21] and the observed instrument line shape, which is about a factor of three shy of the VIPA-limited resolution predicted by using an analytical expressions and assuming several key parameters for the VIPA (see Table 1). A subsequent demonstration of a VIPA spectrometer in the near infrared achieved $RP$ = 230 000 for measurements of gas composition and breath analysis [2], with a design having potential $RP$ = 580 000 based on the detailed consideration of numerous optical components [22]. The first mid-infrared VIPA spectrometer reported $RP$ = 130 000 [23], lower than the VIPA-limited prediction of $RP$ = 390 000. More recently, a far-infrared VIPA was fabricated with measured $RP$ = 16 300 at $\lambda$ = 116 μm and cryogenic temperature ($T$ = 4 K) [24], with a corresponding VIPA-limited $RP$ = 22 000. The key parameters from these initial VIPA spectrometers are summarized in Table 1 to illustrate the challenges in achieving resolution limits when implementing proof-of-principle devices and instruments in new measurement regimes.

Numerical simulations offer a powerful tool to bridge the gap between analytical expressions for VIPA-limited resolving power and experimental observations of newly realized spectrometers. Furthermore, computational models allow for the controlled variation of parameters and components that are difficult or time-consuming to adjust experimentally. Therefore, verified numerical simulations can enable high-performing VIPA spectrometers when utilized early in the design and testing phase. In this context, comprehensive system-level modeling is crucial—not only for design optimization but also for evaluating sensitivity, reducing prototyping, and predicting performance under different operating conditions.

**Table 1. Virtually imaged phased array (VIPA) spectrometers first reported in the visible, near-infrared, mid-infrared, and far-infrared spectral regions. Data includes wavelength ($\lambda$), highly reflective coating power reflection coefficient ($R_1$), partially reflective coating power reflection coefficient ($R_2$), VIPA thickness ($t$), incident beam angle ($\theta_i$), and spacer material refractive index ($n'$). Also listed are the experimental (Exp.) and theoretical (Theo., from the ideal analytical expressions in Section 2) resolving powers ($RP$) and their ratio.**

| Study | $\lambda$ (µm) | $R_1$ (%) | $R_2$ (%) | $t$ (mm) | $\theta_i$ (°) | $n_r$ | Exp. RP | Theo. RP | Exp./Theo. (%) |
|---|---|---|---|---|---|---|---|---|---|
| Diddams *et al.* [20] | 0.633 | 100* | 96 | 2* | 2* | 1.5* | 400 000 | 1 400 000 | 29 |
| Thorpe *et al.* [2,22] | 1.6 | 100 | 96 | 1.5 | 2 | 2 | 230 000 | 580 000 | 40 |
| Nugent-Glandorf *et al.* [23] | 3.8 | 99.8 | 98 | 0.8 | 15 | 3.43 | 130 000 | 390 000 | 33 |
| Zou *et al.* [24]† | 115.7 | 99 | 92 | 9.9 | 13 | 3.39 | 16 300‡ | 22 000 | 74‡ |

*Values for these parameters were not reported in Diddams *et al.* [20]. The values listed here are estimated for a VIPA free spectral range of $\nu_{FSR}$ = 50 GHz, assuming a solid glass spacer and a small incident beam angle.
†Information from Zou *et al.* [24] is taken from their Table 1 and Figure 2 for a VIPA operating at a temperature of 4 K, where VIPA spacer material absorption losses were negligible. At 300 K, a resolving power of 5800 was reported.
‡The experimental results of Zou *et al.* [24] were for a VIPA only, and not for a cross-dispersed spectrometer.

In this work, we performed a systematic analysis to assess the impact of each optical component within a new mid-infrared VIPA spectrometer assembly, aiming to identify the critical factors contributing to instrument line shape broadening and distortion. Using analytical expressions, we calculated an ideal, theoretical resolving power for our VIPA spectrometer at $\lambda$ = 4.6 µm of $RP$ = 830 000. Through numerical simulation, we identified spherical aberration, VIPA height, imaging lens alignment, and aperture effects as the primary sources of resolution loss during initial spectrometer alignment. By addressing some of these initial deficiencies, we demonstrated that the spectrometer resolving power could be as high as $RP$ = 440 000 and thereby gained a better understanding of the discrepancy between analytical expression predictions and experimental results at the component level.

The experimental result is a mid-infrared VIPA spectrometer capable of resolving individual frequency-comb modes separated by a laser repetition rate of $f_{rep}$ = 250 MHz. Without the use of a filter cavity (e.g., [23]), we demonstrated a high-resolution frequency-comb spectrometer at $\lambda$ = 4.53 µm, pushing the prospects of performing precision molecular spectroscopy further into the mid-infrared than previous demonstrations of comb-mode resolved VIPA spectrometers at wavelengths near 3.3 µm and 3.6 µm, respectively [10,25].

The paper is organized as follows. The VIPA-limited resolving power is calculated in Section 2 using analytical expressions to verify idealized numerical simulations and identify component limitations. We report an experimentally measured resolving power for the new cross-dispersed spectrometer in Section 3 using a continuous-wave quantum cascade laser (QCL) source and compare it to numerical simulations performed with realistic optical components and manufacturer-specified parameters. The resolving power is improved in Section 4, approaching the limit predicted by our more realistic numerical simulations. The VIPA angular dispersion is measured in Section 5 and compared with realistic numerical simulations. In Section 6, the VIPA optical throughput is measured and, through numerical simulations, excess losses in the partially reflective coating are correlated with optical throughput. Finally, in Section 7, individual comb modes from a mid-infrared laser frequency comb source with repetition rate $f_{rep}$ = 250 MHz are resolved by the spectrometer.

## 2. VIPA resolving power in the ideal limit

Like a Fabry–Pérot interferometer, a VIPA consists of two reflective surfaces: one with a highly reflective (HR) coating and the other with a partially reflective (PR) coating. Light enters at a non-normal angle of incidence through an anti-reflective (AR) coated window and undergoes multiple displaced internal reflections, forming a virtual curve of sources in front of the VIPA that are imaged onto a detector plane as illustrated in Fig. 1(a). Interference between these virtual sources results in high angular dispersion, often exceeding that of a conventional diffraction grating and making VIPA devices effective elements for high-resolution spectroscopy.

### *2.1 Analytical expressions*

The VIPA was modeled here, prior to assembly of the cross-dispersed spectrometer, using a line-focused continuous-wave laser beam, a spherical imaging lens, and a plane array detector. In Fig. 1(a), a series of red dots show the expected transmission results imaged at the detector plane. The VIPA-limited line shape function is Lorentzian, and the relative intensity of the individual peaks is defined by the incident beam shape function (i.e., Gaussian).

The peak-normalized transmitted intensity profile of the VIPA, $I_{\mathrm{n}}^{\mathrm{t}}$, was calculated in one dimension from the plane wave formalism of Hu et al. [26] and is expressed in Eq. (1).

$$I_{\mathrm{n}}^{\mathrm{t}}(x_{\mathrm{im}}) = \exp\left(-2\frac{f_{\mathrm{cyl}}^2 x_{\mathrm{im}}^2}{W^2 f_{im}^2}\right)\frac{(1-r_1 r_2)^2}{(1-r_1 r_2)^2+4(r_1 r_2)\sin^2(k\Delta/2)} \tag{1}$$

In Eq. (1), $f_{\mathrm{cyl}}$ is the cylindrical lens focal length, $x_{\mathrm{im}}$ is the vertical position at the detector array (normal to the incident beam propagation), $W$ is the input Gaussian beam radius, $f_{\mathrm{im}}$ is the imaging lens focal length, $r_1$ and $r_2$ are the electric-field reflection coefficients for the HR and PR coatings with power reflection coefficients $R_1 = r_1^2$ and $R_2 = r_2^2$, $k = 2\pi/\lambda$ is the wave-vector, $\lambda$ is the wavelength, and $k\Delta$ is the one-dimensional dispersion law. From the one-dimensional dispersion law of Hu et al. [26], the $\Delta$ parameter is defined in Eq. (2).

$$\Delta = 2t\sqrt{n'^2 - n^2\sin^2(\theta_{\mathrm{i}} + \theta_\lambda)} \tag{2}$$

Here, $t$ is the VIPA thickness, $n'$ is the refractive index of the VIPA spacer material, $n$ is the refractive index of the surrounding medium, $\theta_{\mathrm{i}}$ is the incident angle of the beam relative to the VIPA entrance window, and $\theta_\lambda$ is the output dispersion angle. The output dispersion angle is related to a vertical position at the detector array by the expression $\tan(\theta_\lambda) = x_{\mathrm{im}}/f_{\mathrm{im}}$. The incident angle, $\theta_{\mathrm{i}}$, is related to the internal VIPA reflection angle, $\theta_{\mathrm{in}}$, by the expression $n\sin(\theta_{\mathrm{i}}) = n'\sin(\theta_{\mathrm{in}})$.

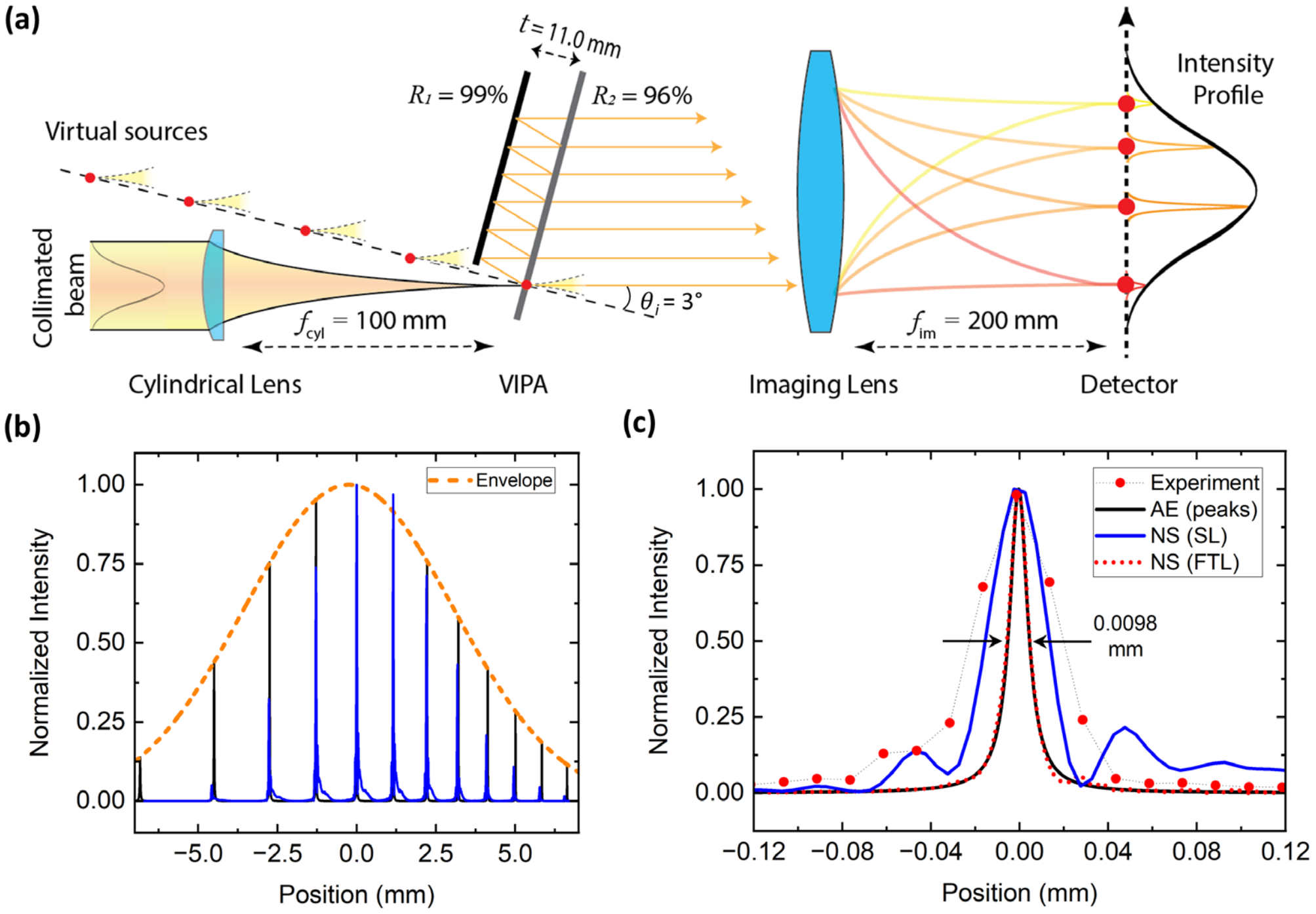

**Figure 1.** (a) Schematic of the virtually imaged phased array (VIPA) and imaging system. (b) Comparison of the VIPA transmission calculated from analytical expressions (black line) and numerical simulations (blue line). (c) Magnified view of a single transmission peak, showing results from experimental data (red dots, see Sections 3 and 4), the analytical expressions (AE, black line), and numerical simulations (NS) imaged with either a spherical lens of 50.8 mm diameter (SL, blue line) or a Fourier-transform lens (FTL, red dotted line).

We identify Eq. (1) as an ideal analytical expression for the transmitted VIPA intensity profile, while noting several key assumptions here. Absorption losses in the VIPA spacer material, as well as absorption and scattering losses in the reflective coatings, were assumed negligible. Polarization and wavelength dependence of the power reflection coefficients were also not considered. The small angle approximation was applied, resulting in the ratio $x_{\mathrm{im}}^2/f_{\mathrm{im}}^2$ appearing here in Eq. (1) instead of the $\sin^2(\theta_\lambda)$ term found in Eq. (10) of Hu et al. [26]. We correct an error in Eq. (10) of Hu et al. [26], using the quantity $W^2$ in Eq. (1) to define the Gaussian profile of the input beam instead of the quantity $W$. Finally, we define ideal as the limit where the integer number of internal reflections, $M$, due to the finite height of the VIPA, $l$, is sufficiently large so that $M \approx l/[2t\tan(\theta_{\mathrm{in}})]$ is greater than the quantity $\left(\pi\sqrt{F_1}\right)/2$, where $F_1 = (4r_1r_2)/(1-r_1r_2)^2$ and $\mathcal{F} = \left(\pi\sqrt{F_1}\right)/2$ is the VIPA finesse.

In the limit where $M > \mathcal{F}$, it is straightforward to show numerically that the broadening terms in Eq. (10) of Hu et al. [26] and our Eq. (1) approach the peak-normalized Airy formula.

$$\frac{1+F_M\sin^2(Mk\Delta/2)}{1+F_1\sin^2(k\Delta/2)} \approx \frac{(1-r_1r_2)^2}{(1-r_1r_2)^2+4r_1r_2\sin^2(k\Delta/2)} \tag{3}$$

In Eq. (3), $F_M = (4r_1^M r_2^M)/(1-r_1^M r_2^M)^2$. The limiting case where $M > \mathcal{F}$ is discussed in more detail in Section 7.6, titled "Multiple-beam interference", of Born and Wolf [27], and by Hu et al. [26] in their derivation of resolving power around Eq. (16). With reference to Born and Wolf [27], Hu et al. [26] state at the end of their Section 2.1 titled "Working principle of VIPA" that "... the VIPA can be treated as a classical optical multiple beam interference device, more specifically, a Lummer-Gehrcke plate." For comparisons between the plane wave dispersion law used here and other proposed VIPA dispersion laws like Vega et. al. [28] and Xiao et al. [29,30], see Hu et al. [26].

The mid-infrared VIPA tested here comprised highly and partially reflective coatings, with manufacturer-specified power reflection coefficients $R_1$ = 99 % and $R_2$ = 96 % separated by a calcium fluoride ($CaF_2$) spacer ($n'$ = 1.4036 at $\lambda$ = 4.6 μm [31]) with a thickness of $t$ = 11.0 mm and height of $l$ = 76.2 mm. For evaluation by analytical expressions, we test the VIPA using a collimated Gaussian input beam with radius $W$ = 3 mm, at a tilt angle of $\theta_i$ = 3 °, and with cylindrical and imaging lens focal lengths of 100 mm and 200 mm, respectively. The black line in Figure 1(b) shows the transmitted intensity profile for the new mid-infrared VIPA, as calculated by Eqs. (1)–(2).

In the ideal limit, the VIPA free spectral range was $\nu_{\mathrm{FSR}} = c/\Delta$ = 9.7 GHz, calculated at $\lambda$ = 4.6 μm and $\theta_\lambda$ = 0 °. The ideal resolving power, defined in terms of the free spectral range as $RP = \mathcal{F}(\nu/\nu_{\mathrm{fsr}})$, was $RP$ = 830 000. The same resolving power was also estimated from the analytical expression by converting the calculated linewidth of 0.0098 mm (annotated in Fig. 1(c)) to optical frequency using the factor of 1.22 mm per free spectral range (i.e., the average peak spacing near $x_{\mathrm{im}}$ = 0 mm in Fig. 1(b)). The same approach—converting simulated linewidth to frequency using the free spectral range—was used to evaluate resolving power for the numerical simulations in Section 2.2.

For a predicted resolving power of $RP$ = 830 000 at $\lambda$ = 4.6 μm, the new mid-infrared VIPA spectrometer would resolve peaks spaced as closely as $\delta\nu$ = 78 MHz, using the FWHM definition of resolving power described by Robertson [21].

### *2.2 Numerical simulations*

To obtain a more realistic estimate of resolving power, numerical simulations were performed for the actual geometry and optical parameters of the cylindrical lens, VIPA, and imaging lens system used. For this purpose, a non-sequential ray tracing approach was implemented in Ansys Zemax OpticStudio [32] to simulate complex light propagation, including multiple reflections and beam overlap within the VIPA and imaging system. As a preliminary step, our numerical simulations were tested against the results of Yang et al. [33]. We also note that optical design principles and software were used recently to evaluate near-infrared spectrograph concepts for exoplanetary science [34,35].

For our new mid-infrared spectrometer, optical components such as lenses were modeled using manufacturer-provided specifications, while the VIPA was built by pairing highly and partially reflective surface separated by a $CaF_2$ spacer. To focus our initial numerical simulations on providing estimates of resolving power, absorption and scattering losses in the VIPA coatings were neglected. Simulations were performed using one million rays, and the detector was placed at the focal plane of the imaging lens. The position of the detector was adjusted to minimize the VIPA transmission linewidth. The simulated detector array matched the infrared camera used in the experiment, with an array of 640 pixels by 512 pixels and a pixel pitch of 15 μm.

Compared in Fig. 1(b) are calculations made using the analytical expression and the results of the numerical simulations, with a zoomed-in trace shown in Fig. 1(c). For comparison in Fig. 1(c), an experimental measurement of transmission peaks from the VIPA output is also shown (see Sections 3 and 4). The analytical expression (solid black line) represented the ideal case and produced the narrowest FWHM and thus defined the theoretical VIPA-limited resolving power. In contrast, the ray-tracing numerical simulation (solid blue line) yielded broader transmission peaks.

The discrepancy between the analytical expressions and numerical simulations is understood by systematically varying the optical component parameters within the numerical simulation. The dominant factors found to influence the resolving power in the numerical simulations were 1) the VIPA height, 2) aberrations introduced by the imaging lens, 3) aperture effects, and 4) the positioning of the imaging lens relative to the VIPA. To isolate the VIPA, several of these limitations were removed by replacing the imaging lens with a Fourier-transform operation applied to the complex-valued electric-field transmitted by the VIPA. This operation likely eliminated the influence of small-angle or paraxial lens approximations found in the ray-tracing software application, and more accurately treated the large angular dispersion characteristics of the VIPA device by emulating an aberration-free imaging lens.

As shown in Fig. 1(c) by the red dotted line, the Fourier transform lens (FTL) approach improved agreement between the numerical simulation and analytical expression results. This is also illustrated in Table 2, where resolving powers are compared between analytical expressions and numerical simulations. For Table 2, values of $\lambda$ = 4.6 µm, $f_{cyl}$ = 100 mm, $f_{im}$ = 200 mm, $\theta_i$ = 3 °, $R_1$ = 0.99, and $R_2$ = 0.96 were fixed, and the analytical expressions (AE) again predicted a value of $RP$ = 830 000. Numerical simulations (NS) performed for a hypothetically long VIPA ($l$ = 177.8 mm) and the FTL approach showed a slightly reduced $RP$ = 700 000, which we attribute to near-field effects resulting from the large VIPA aperture. When the VIPA height was reduced to match the experimental value of $l$ = 76.2 mm, a further reduction to $RP$ = 610 000 was observed. Finally, Table 2 reveals additional limitations due to spherical lens (SL) aperture and aberration which reduced achievable resolving power to $RP$ = 580 000 and $RP$ = 400 000, respectively. Comparably, the initial experimental results discussed in Section 3 achieved a VIPA resolving power of $RP$ = 290 000.

**Table 2. Comparison of VIPA resolving powers determined from analytical expressions (AE), numerical simulations (NS), and experiment (Exp.). Common parameters were $\lambda$ = 4.6 µm, $f_{cyl}$ = 100 mm, $\theta_i$ = 3 °, $R_1$ = 0.99, $R_2$ = 0.96, and $f_{im}$ = 200 mm. Fourier transform lens (FTL) and spherical lens (SL) imaging lens types were tested in numerical simulations at a fixed VIPA-to-lens distance of 1 m. Proposed limitations listed in the final column accumulated in subsequent rows.**

| Method | VIPA height (mm) | Imaging lens type | Imaging lens diameter (mm) | Resolving power | Proposed limitation |
|---|---|---|---|---|---|
| AE | | | | 830 000 | Ideal limit |
| NS | 177.8 | FTL | 177.8 | 700 000 | Near-field effects |
| NS | 76.2 | FTL | 76.2 | 610 000 | VIPA height |
| NS | 76.2 | FTL | 50.8 | 580 000 | Imaging lens aperture |
| NS | 76.2 | SL | 50.8 | 400 000 | Imaging lens spherical aberration |
| Exp. | 76.2 | SL | 50.8 | 290 000 | |

The numerical simulations suggested that increasing the VIPA height and employing aberration-free lens would significantly enhance spectrometer performance, thereby reducing the gap between VIPA-limited predictions from analytical expressions and experiment. As described in Section 2.1, the minimum VIPA height required to approach the resolving power limit set by the ideal analytical expressions depends strongly on the reflectivity of the HR and PR surfaces, spacer absorption, and the angle of incidence [26,27]. In the next section, the numerical simulations are extended to the full cross-dispersed spectrometer assembly and are again used to explore the key parameters responsible for deviations from performance limits.

## 3. Cross-dispersed spectrometer: Initial experimental results

The experimental setup and simulation layout are shown in Fig. 2(a,b). The laser was launched from a fiber collimator with free-space output beam radius of $W$ = 3 mm and line-focused by a cylindrical lens of $f_{cyl}$ = 100 mm. The VIPA specifications were previously listed in Section 2.1 and in Fig. 1. A diffraction grating with a groove number of 300

mm$^{-1}$ received the VIPA output, and the first-order output from the diffraction grating was steered by a flat mirror to the imaging lens with $f_{\mathrm{im}}$ = 200 mm, and then ultimately onto a detector array and camera (size and pixel pitch mentioned in Section 2.2). The spectral response of the camera was from $\lambda$ = 1.5 µm to $\lambda$ = 5.1 µm, and the integration time and frame rate were set to 500 µs and 100 Hz.

The detector images for a laser wavelength of $\lambda$ = 4.6 µm are shown in Fig. 2(c,d) for both the experimental measurement and the numerical simulation. The corresponding cross-sectional line profiles of the peaks are presented in Fig. 2(e,g), while a magnified view of the central peaks fitted to Lorentzian functions are provided in Fig. 2(f,h). (Note that fitting to more advanced line shape functions [36], including convolutions of multiple functions [37], is not the focus of this work.) From these Lorentzian fits, the FWHM values were 2.8 pixels for the experiment and 2.0 pixels for the simulation. In Fig. 2(e,g), the peak spacing spanned approximately 120 pixels, sufficient space to accommodate about 43 (experiment) and 60 (simulation) peaks within one free spectral range. The corresponding resolving power is therefore estimated to be $RP$ = 290 000 (experiment) and $RP$ = 400 000 (numerical simulation).

## 4. Cross-dispersed spectrometer: Experimental optimization

Experimentally, the imaging lens and grating height were adjusted to maximize their collection of light from the VIPA. To mitigate spherical aberration, a lens with a longer focal length was used. Since a longer focal length corresponds to a larger radius of curvature at a fixed lens diameter, aberrations were reduced when the lens f-number was increased (where the lens f-number is $f/\# = f_{\mathrm{im}}/D$, and $D$ is the lens diameter). To demonstrate this effect, we compared experimental results obtained using an imaging lens with $f_{\mathrm{im}}$ = 1000 mm ($f/\#$ = 20) to another with $f_{\mathrm{im}}$ = 200 mm ($f/\#$ = 4). As shown in Fig. 3, the resolving powers achieved with the 200 mm and 1000 mm lenses are $RP$ = 290 000 and $RP$ = 440 000, respectively. The longer focal length imaging lens also made better use of the detector array by magnifying the image and stretching one free spectral range over more pixels. These results underscore the critical role of the imaging lens in achieving VIPA-limited spectrometer performance and suggested that an aberration-free imaging lens with a large f-number would enable enhanced resolving power.

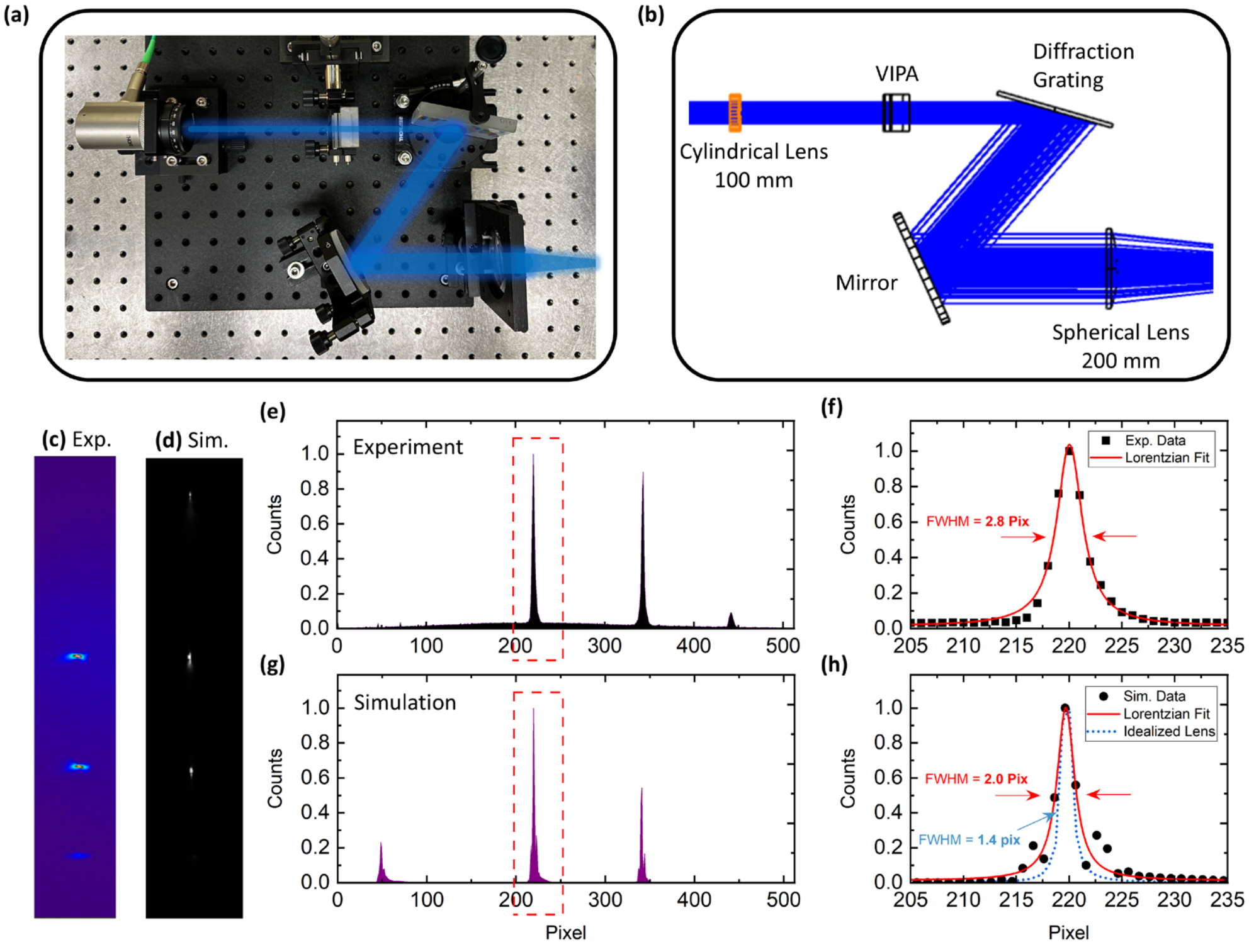

**Figure 2.** (a) Experimental setup and (b) layout for numerical simulation. (c,d) Detector images from experiment and simulation. (e,g) Cross-sectional profiles of the 2D images. (f,h) Zoomed-in view of the central peak with the corresponding Lorentzian fit (red lines).

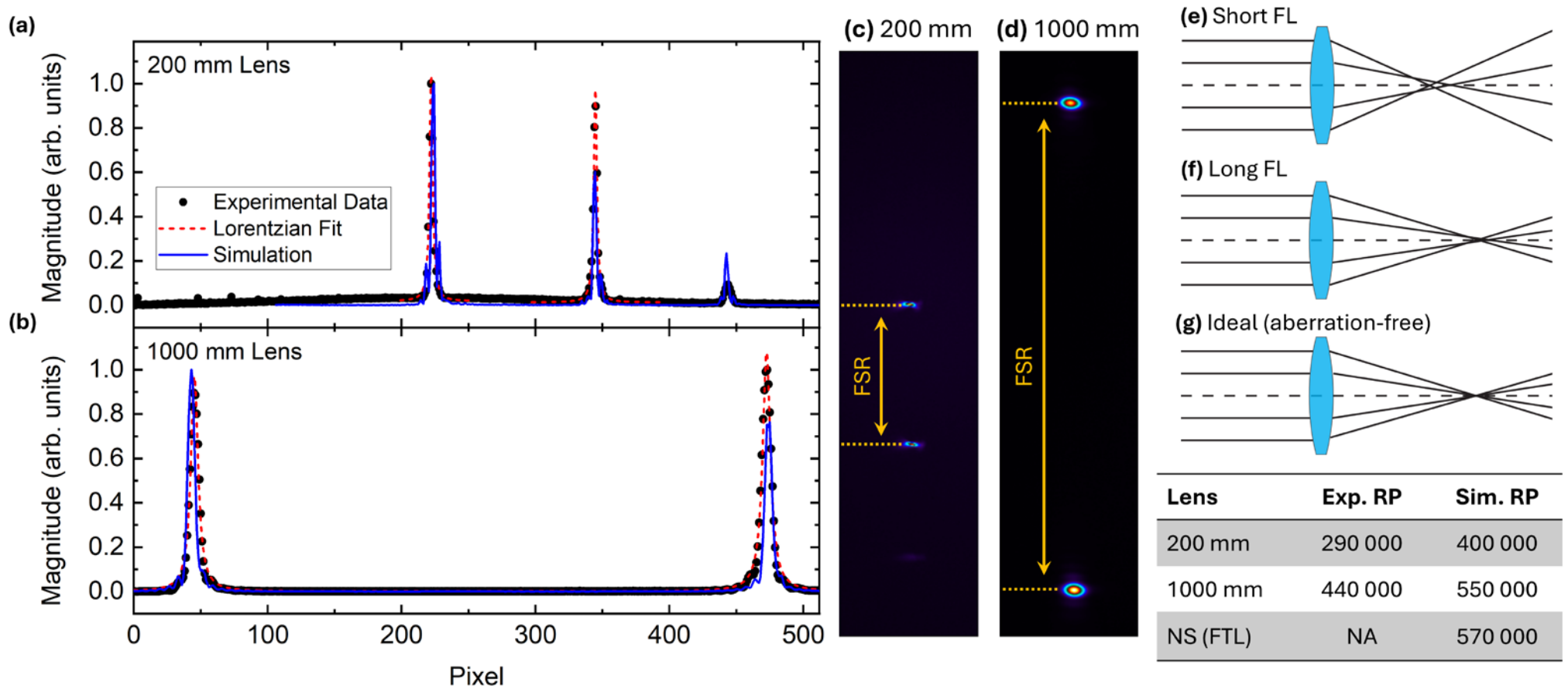

**Figure 3.** The effect of imaging lens focal length at fixed lens diameter on the VIPA resolving power. Numerical simulations (blue lines) were performed with (a) $f_{\mathrm{im}}$ = 200 mm and (b) $f_{\mathrm{im}}$ = 1000 mm imaging lenses and are compared to experimental results (red dashed lines). VIPA transmission is shown for two experimental lens focal lengths, (c) $f_{\mathrm{im}}$ = 200 mm and (d) $f_{\mathrm{im}}$ = 1000 mm. (e–g) Schematics of spherical aberration from short focal-length, long focal-length, and ideal aberration-free lenses. The experimental resolving powers achieved are listed in the table in the image, along with the corresponding ideal value from numerical simulations with Fourier transform lens (FTL).

## 5. VIPA angular dispersion

Angular dispersion determines how well different frequencies are resolved at the detector. A schematic of angular dispersion is presented in Fig. 4(a), where each peak corresponds to a distinct wavelength. Here, the laser was tuned by adjusting the QCL temperature, with a tuning coefficient ≈ 0.43 nm $\mathrm{K}^{-1}$ determined via curve fitting of the transmission from a commercial silicon etalon. As the wavelength was adjusted, the position of the transmission peak shifted across the detector plane.

Figure 4(a)i–iv shows camera images captured as the QCL wavelength was incrementally increased from $\lambda$ = 4600.0 nm to $\lambda$ = 4602.5 nm. Due to the VIPA angular dispersion, increases in the laser wavelength manifested as changes in the dispersion output angle and the motion of the transmission peaks in the plane of the detector array. The output angle ($\theta_\lambda$) was estimated from the vertical displacements of the peak positions ($x_{\mathrm{im}}$) on the image plane and the focal length of the imaging lens ($f_{\mathrm{im}}$) using the small angle approximation $\theta_\lambda \approx x_{\mathrm{im}}/f_{\mathrm{im}}$. The black circles in Fig. 4(b) represent the experimentally measured VIPA output angle as a function of laser wavelength, while the open circles show the corresponding results from the numerical simulations. For these measurements, the incident angle of the VIPA was set to $\theta_{\mathrm{i}} \approx 1.5$ °. From the plane wave formalism of Hu et al. [26], the wavelength dispersion was related to the output angle via Eq. (4).

$$\Delta\lambda = \lambda_p - \lambda_0 = \frac{\lambda_0}{n' \cos(\theta_{\mathrm{in}})}\left[\sqrt{n'^2 - n^2 \sin^2(\theta_{\mathrm{i}} + \theta_\lambda)} - \sqrt{n'^2 - n^2 \sin^2(\theta_{\mathrm{i}})}\right] \quad (4)$$

The vertical separation between curves observed in Fig. 4(b) corresponded to the VIPA free spectral range. Over one free spectral range, the output angle varied approximately from −0.5 ° to 0.5 °, yielding an angular dispersion of ≈ 1.2 ° $\mathrm{nm}^{-1}$. Compared to calculations from analytical expressions, where $\lambda_{\mathrm{FSR}}$ = 0.69 nm, we determined from fits to Eq. (4) an $\lambda_{\mathrm{FSR}}$ = 0.67 nm from the experimental data (solid lines) and $\lambda_{\mathrm{FSR}}$ = 0.66 nm from the numerical simulation (dashed lines). The slight mismatch between experimental results and numerical simulations near $\theta_\lambda = 0$ ° is attributed to small experimental uncertainties in setting the incident angle to $\theta_{\mathrm{i}}$ = 1.5 °, defining the physical location of $\theta_\lambda$ = 0 in the laboratory frame, and the laser tuning curve and wavelength calibration.

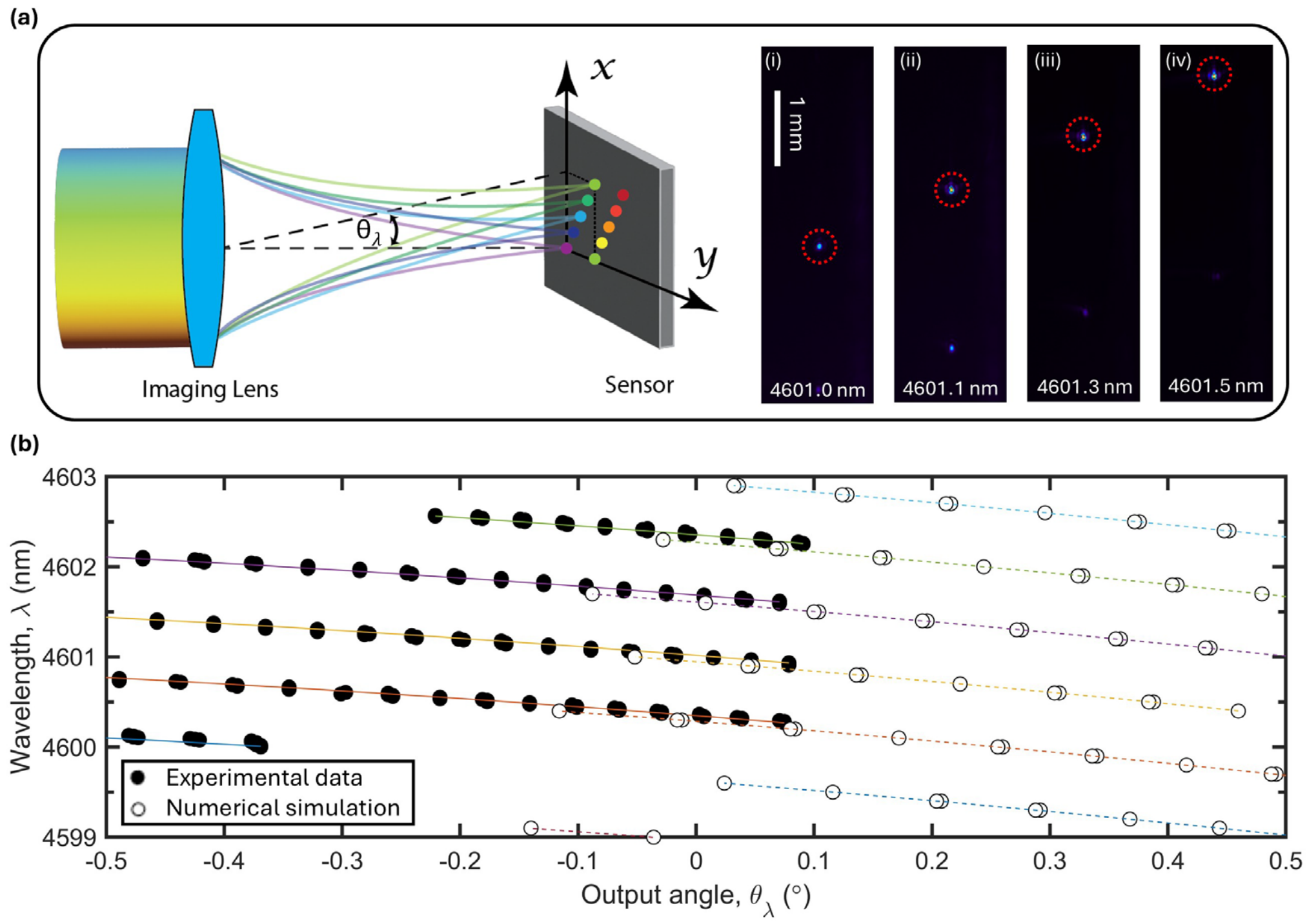

**Figure 4.** (a) Schematic of the angular dispersion measurement, along with the detector image captured for wavelengths ranging from (i) 4601.0 nm to (iv) 4601.5 nm. (b) Experimental and numerical simulation results show the laser wavelength as a function of the VIPA output angle, overlaid with the corresponding fits based on Eq. (4). Solid lines are fitted to the experimental data, and dashed lines are fitted to the numerical simulation.

## 6. VIPA optical throughput

The numerical simulations enabled an evaluation of the VIPA optical throughput, a value constrained by the VIPA coating reflectivity, transmission, and absorption plus scattering (excess) losses. Figure 5 presents the numerically simulated optical throughput of the VIPA as a function of coating absorption losses at the PR surface (assuming zero scattering), with the reflectivity of the mirrors held constant and assuming no excess losses at the AR-coated input window or the HR coating. The results revealed a linear trend, where increased absorption losses at the PR face resulted in decreased optical throughput.

We experimentally measured the optical power throughput of the VIPA using a continuous-wave laser with wavelength $\lambda = 4.6$ µm. The input power ($P_{in}$) was measured directly before the VIPA, and a lens with focal length of 200 mm was used to collect most of the transmitted laser light and focus it onto a thermal power meter ($P_{out}$). Although the lens with diameter of 50.8 mm did not capture rays from very high bounce numbers near the top of the 76.2 mm tall VIPA, the numerical simulations suggest that the lens did capture most of the transmitted optical power. The measured optical power throughput, defined as $P_{out}/P_{in}$, was 60 %.

As shown in Fig. 5, the numerically simulated ideal case—where the PR surface has no absorption and 4 % transmission (96 % reflectivity)—yielded a throughput near 75 %. When the PR transmission was reduced to 2 % due to absorption or scattering losses, the throughput dropped sharply to below 50 %. By comparing the measured results with numerical simulations, we estimated that the PR surface transmission was approximately 3.2 % and the absorption was about 0.8 %. These results highlight the critical importance of minimizing absorption and employing high-quality, low-loss coatings in VIPA construction to maximize optical throughput.

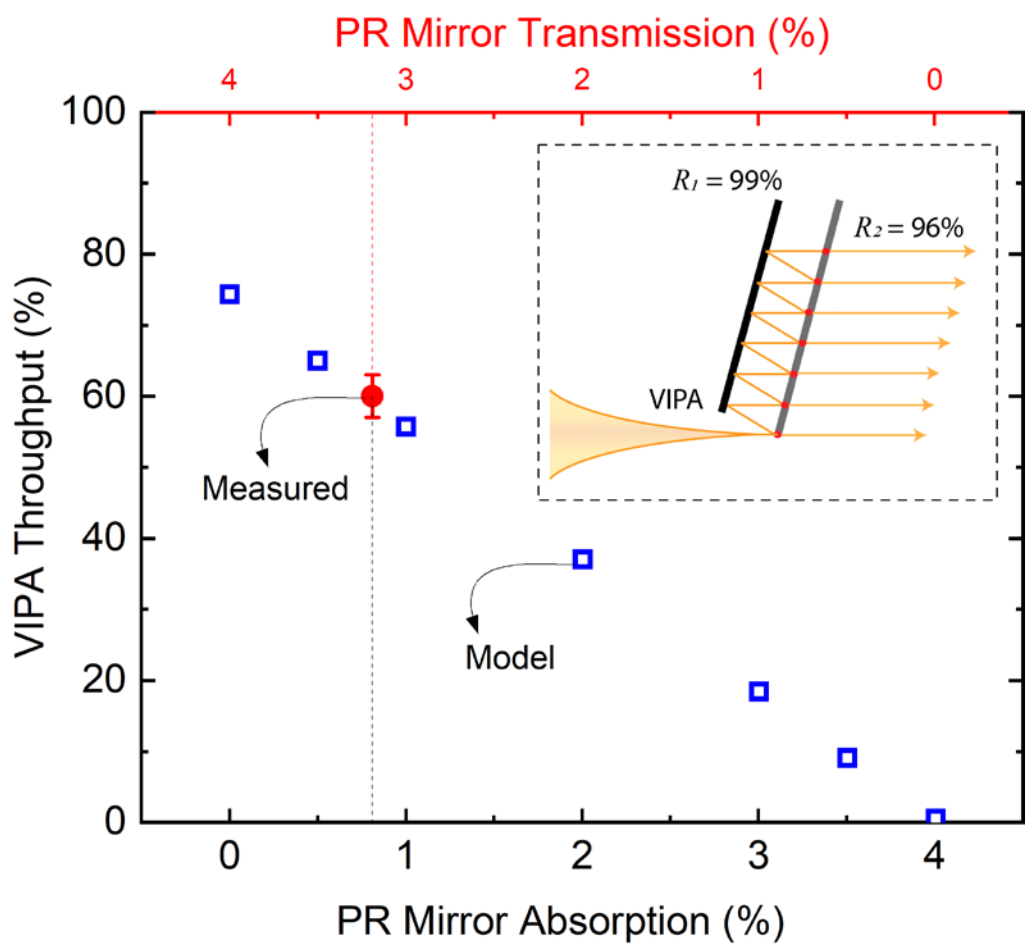

**Figure 5.** VIPA optical throughput ($P_{out}/P_{in}$) at λ = 4.6 µm as a function of absorption in the partially reflective (PR) coating. The measured value for the device studied here is shown as a solid red circle, with error bars representing the standard deviation across four independent measurements. The transmission and absorption were not directly measured but were instead inferred from the numerical simulation. The reflectivities at both the HR and PR coatings were kept constant in the simulation, while only the absorption of the PR surface was varied.

## 7. Laser frequency comb resolved

This section describes the use of a laser frequency comb source with a repetition rate of $f_{rep}$ = 250 MHz to test the broadband resolving capabilities of our new VIPA spectrometer. At a wavelength of $\lambda$ = 4.53 µm, the comb source mode-spacing corresponded to approximately 17 pm, which required a FWHM resolving power of at least 260 000. The previous Section 4 showed that the system—with the 1000 mm focal length imaging lens—would reach resolving powers as high as 440 000, suggesting that the spectrometer should resolve adjacent comb modes.

Figure 6 shows a comparison between experimental measurements and numerical simulations of the VIPA spectrometer illuminated by the laser frequency comb source. The experimental two-dimensional spectrogram captured by the mid-infrared camera is shown in Fig. 6(a), while the numerical simulation is shown in Fig. 6(b). The bright dot visible in the top and the bottom of the central portion of Fig. 6(a) is a continuous-wave laser co-propagated with the comb source, illustrating that one free spectral range is imaged by the spectrometer. Individual comb modes are clearly resolved along the vertical VIPA dispersion dimension, highlighted by a dashed, overlaid box. The box highlights a portion of the image which is further illustrated in Fig. 6(c). In Fig. 6(c), the comb modes appear as a series of peaks separated by an average value of about 11 pixels and in good agreement with the spacing observed in the numerical simulation of Fig. 6(b).

## 8. Conclusion

In this work, we used numerical simulations to understand discrepancies between analytically predicted and experimentally observed resolving powers in a virtually imaged phased array (VIPA) spectrometer. While the analytical expressions for the mid-infrared VIPA predicted a resolving power of *RP* = 830 000 at a wavelength of $\lambda$ = 4.6 µm, the initial measurements yielded lower values by about a factor of three. Through systematic analysis, the VIPA height and spherical aberration from the imaging lens were identified as the dominant limiting factors. By employing a longer focal length imaging lens, the resolving power more than doubled to *RP* = 440 000, and numerical simulations suggested values higher than *RP* = 570 000 would be possible with aberration-free optics and a taller VIPA structure. For the complete cross-dispersed spectrometer, the achieved resolving power corresponded to 80 % of the limit determined by numerical simulations.

The numerical simulations succeeded in providing a realistic model for comparison with experimental results. However, the influence of fabrication uncertainties and alignment tolerances was not considered here. Uncertainties in the VIPA specifications like thickness and coating reflectivity, as well as uncertainty in the experimental alignment for parameters like angle of incidence, will also influence how closely we perceive our experimental results as agreeing with numerical simulations or analytical expressions. A full evaluation of experimental uncertainties in VIPA spectrometers would therefore provide greater confidence across all comparisons.

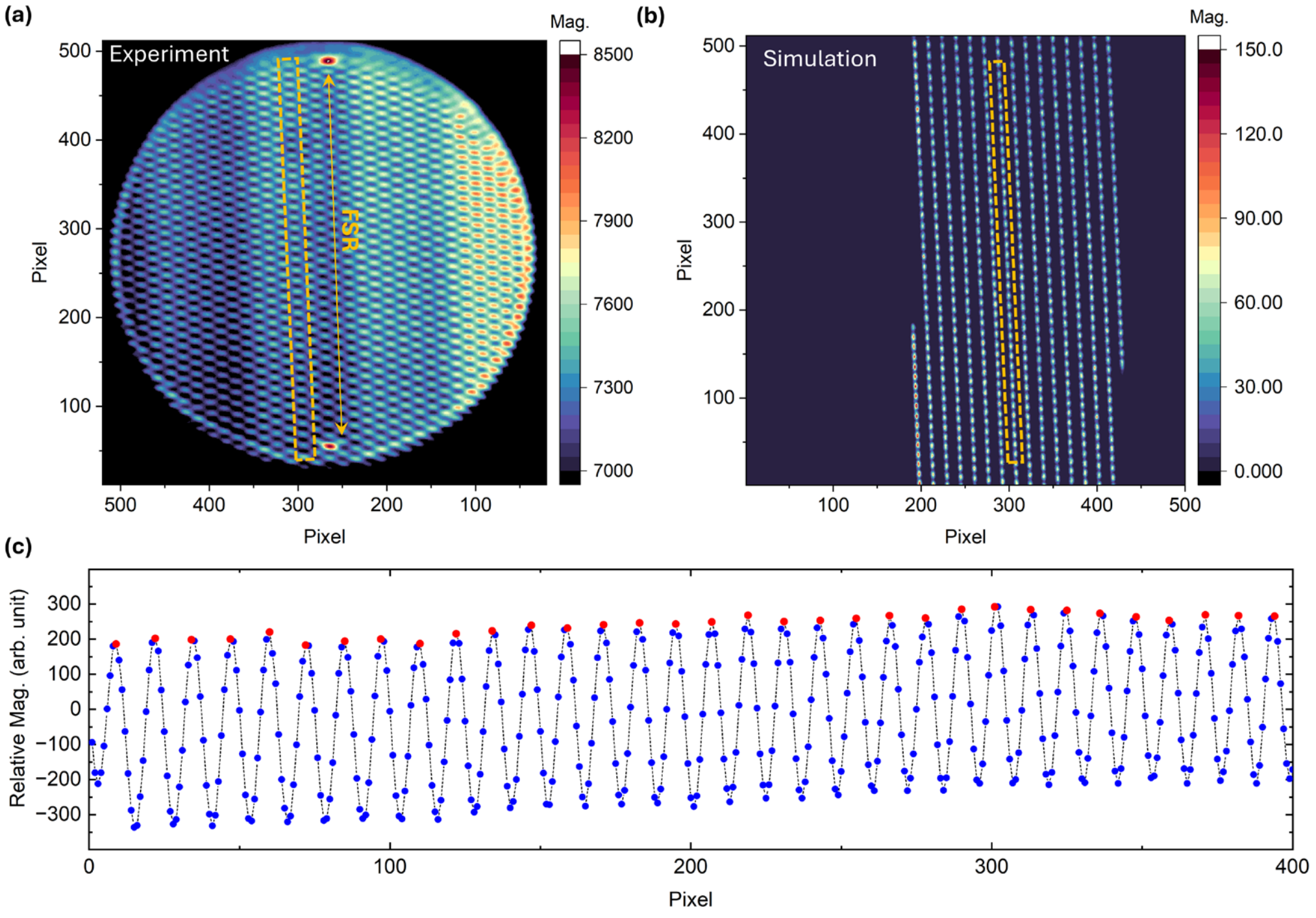

**Figure 6.** The mid-infrared VIPA spectrometer illuminated by a laser frequency comb. (a) Experimental spectrogram with co-propagated continuous-wave laser. (b) Numerical simulation. (c) Experimental vertical line-out from the dashed-box region of (a) is shown as a blue dashed line and blue dots. Individual comb-mode peaks are highlighted by red dots.

**Funding.** KA; NIST PREP Cooperative agreement 015276772. SIW, AJF; NIST Innovation in Mission-driven Science (IMS) Program. AJF; CHIPS Metrology Program, part of CHIPS for America, NIST, U.S. Department of Commerce. DMB, SIW, AJF; NIST.

**Acknowledgment.** Eisen Gross (NIST) and Kevin Cossel (NIST) for commenting on the paper.

**Disclosures.** The authors declare no conflicts of interest.